\documentclass[fleqn,aps,pra,amsmath,amssymb,reprint,showpacs]{revtex4-1}

\usepackage{amsmath}
\usepackage{graphicx}
\usepackage{dcolumn}
\usepackage{bm}
\usepackage{textcomp}

\def\ket#1{\mathinner{|{#1}\rangle}}

\begin{document}

\title{Observation of coherent population trapping in a V-type two-electron system}
 \author{Alok K. Singh}
 \affiliation{Department of Physics, Indian Institute of
 Science, Bangalore 560\,012, India}
 \author{Vasant Natarajan}
 \affiliation{Department of Physics, Indian Institute of
 Science, Bangalore 560\,012, India}
 \email{vasant@physics.iisc.ernet.in}
 \homepage{www.physics.iisc.ernet.in/~vasant}

\begin{abstract}
We observe coherent population trapping (CPT) in a two-electron atom---$^{174}$Yb---using the $\rm {^1S_0,F=0} \rightarrow {^3P_1,F'=1}$ transition. CPT is not possible for such a transition according to one-electron theory because the magnetic sublevels form a V-type system, but in a two-electron atom like Yb, the interaction of the electrons transforms the level structure into a $ \Lambda $-type system, which allows the formation of a dark state and hence the observation of CPT. Since the two levels involved are degenerate, we use a magnetic field to lift the degeneracy. The single fluorescence dip then splits into five dips---the central unshifted one corresponds to coherent population oscillation, while the outer four are due to CPT. The linewidth of the CPT resonance is about 300 kHz and is limited by the natural linewidth of the excited state, which is to be expected because the excited state is involved in the formation of the dark state. \\

\end{abstract}


\maketitle

\section{Introduction}
Coherent population trapping (CPT) is a widely studied phenomenon since its first observation in 1976 \cite{AGM76}, because of its potential applications in fields like atomic clocks \cite{BNW97}, electromagnetically induced transparency (EIT) \cite{HAR97}, lasing without inversion \cite{MOC00}, stopping and storing of light pulses \cite{HHD99}, optical memory and quantum computing \cite{KRS01}, precise magnetometry based on the nonlinear Faraday effect \cite{NOW02}, and so on.

As the name implies, CPT is a phenomenon in which the atomic population is driven into a dark non-absorbing state in the presence of two {\em coherent} laser beams---usually called the pump and the probe. The population remains trapped there until some decoherent process like spontaneous emission takes place. Therefore, CPT requires such a dark state to be formed, which happens in $\Lambda$-type three-level systems where there are two long-lived lower levels. By the same token, CPT does not occur in V-type systems where the presence of two short-lived upper levels prevents the formation of a dark state. 

Many CPT experiments have been done in \textit{degenerate} two-level systems, where the magnetic sublevels of the levels involved form the required $\Lambda$-type system. For example, CPT has been observed on a $1 \rightarrow 0 $ transition of the $^{87}$Rb D$_2$ line \cite{ACD01}, where the pump and probe lasers have opposite circular polarizations and therefore couple sublevels with selection rules $\Delta m = \pm 1$. In the same manner, CPT is not possible for a $0 \rightarrow 1 $ transition, because the magnetic sublevels can only form a V-type system. 

Here, we report the first observation of CPT for a $0 \rightarrow 1 $ transition, made possible because it is a two-electron atom. The atom used is $^{174}$Yb (with nuclear spin 0), and the transition used is the $\rm {^1S_0} \rightarrow {^3P_1}$ intercombination line at 556 nm. It has been shown theoretically \cite{MCR01} that the interaction of the two electrons combined with the Pauli exclusion principle transforms the V-type system for the two-electron levels into an equivalent $\Lambda$-type system for the one-electron levels. The only conditions are: (i) that the two electrons in the upper levels have parallel spins ({\em orthosystem} in the language of He), and (ii) that the upper levels have opposite parity from that of the lower level. Both these conditions are satisfied for the $\rm {^1S_0} \rightarrow {^3P_1}$ transition used in this study.

Since the above transition is a two-level system, the related phenomenon of coherent population oscillation (CPO) also becomes relevant. CPO is the effect where the same two levels are coupled by the pump and probe beams \cite{BLB03a}, and hence the population shows periodic modulation at the beat frequency of the two fields. CPO has important applications similar to those of CPT \cite{LKB12,KLA13,MBG14}, but is more advantageous because (i) there is no third level involved, so that it can be potentially observed in a larger class of materials including gases, liquids, and room-temperature solids, and (ii) it does not require a two-photon resonance, and therefore laser jitter becomes irrelevant.

One way to separate the two effects is to use a magnetic field. This will cause the CPT resonances to shift, while the CPO resonance will not. In our experiment, we use a magnetic field of 330 mG, which causes the spectrum to split into five equispaced fluorescence dips, spaced apart by about 700 kHz. The central unshifted one corresponds to CPO, while the four outer ones are due to CPT. The observed linewidth is about 300 kHz, limited by the 185 kHz natural linewidth of the excited $\rm ^3P_1$ state, which is expected because the dark state involves the upper state. This linewidth is larger than what is seen in normal CPT resonances of one-electron atoms with $\Lambda$-type level structure, where the dark state involves two lower levels and is therefore extremely narrow. 

\section{Experimental details}
\begin{figure}
\centering{\resizebox{0.95\columnwidth}{!}{\includegraphics{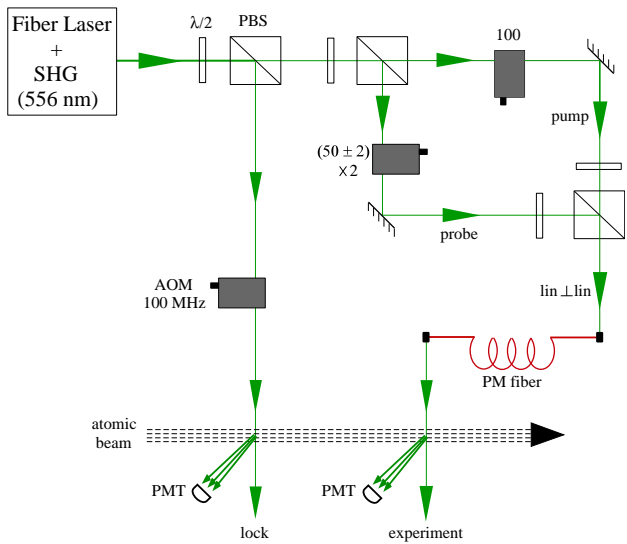}}}
\caption{(Color online) Schematic of the experiment. The probe beam is scanned by $\pm 4$ MHz using a double-passed AOM. Figure key: $\lambda/2$ -- halfwave retardation plate; PBS -- polarizing beam splitter; PM fiber -- polarizing maintaining fiber; PMT -- photomultiplier tube.}
 \label{exptsetup}
\end{figure}

The main parts of the experimental setup are shown schematically in Fig.\ \ref{exptsetup}. The atomic beam is present inside an ultra-high vacuum (UHV) chamber, which is used for both locking the laser and for the spectroscopy. It is generated by resistively heating a quartz ampoule containing unenriched Yb to a temperature of about $400^\circ$C. The source part is connected to the main experimental chamber consisting of a tube with 39~mm diameter $\times$ $125$~mm length, sandwiched between ports for optical access. The entire chamber is maintained at a pressure below $10^{-8}$ torr by a $20$~l/s ion pump.

The $\rm {^1S_0} \rightarrow {^3P_1}$ transition used in this study is at 556 nm. It is generated by frequency doubling the output of a fiber laser. The fiber laser (Koheras Boostik Y10) operates at 1111 nm, and has an output power of $0.5$ W with a linewidth of 70 kHz. Its frequency is doubled to 556 nm in an external cavity doubler (Toptica Photonics) using potassium niobate as the nonlinear crystal to give a total power of 65 mW. As shown in the figure, part of this beam is used for locking the laser to the correct transition. For this, the laser beam is sent perpendicular to the atomic beam (to minimize the first-order Doppler effect), and the resulting fluorescence is collected using a photomultiplier tube (PMT) from Hamamatsu (R928). The locking is achieved by (i) passing the beam through an acousto-optic modulator (AOM) with an up-shift of 100 MHz, (ii) frequency modulation of the AOM shift at 20 kHz, and (iii) lock-in detection of the PMT fluorescence signal to generate the error signal, which is fed back to a piezoelectric transducer that determines the fiber laser frequency.

The remaining part of the laser beam is further divided into two parts---one as a probe beam and other as a pump beam---using a polarizing beam splitter cube (PBS). The two beams therefore have orthogonal linear polarizations. The pump beam is fixed in frequency. By up-shifting by 100 MHz using a second AOM, its frequency is kept the same as that of the laser and is resonant with the transition. The probe beam is scanned around this transition by double passing through a third AOM at $50 \pm 2$ MHz. The double passing ensures directional stability as the AOM is scanned, and the variable shift of $\pm 2$ MHz allows a scan range of up to 8 MHz around the resonance. The two beams (with orthogonal polarizations) are combined on another PBS, and then transported to the experimental chamber through a polarizing maintaining (PM) fiber. There are two advantages to using the PM fiber. One is that it ensures that both beams are perfectly overlapped, and propagate in the same direction as the probe beam is scanned. The second is that the mode shape coming out is perfectly Gaussian. The power in each beam is 40 \textmu W and its $1/e^2$ diameter is 5 mm, hence the intensity at beam center is 0.41 mW/cm$^2$---using the saturation intensity of 0.14 mW/cm$^2$ this gives a Rabi frequency of $1.2 \, \Gamma$.  The output of the fiber is sent across the atomic beam, and the fluorescence signal detected using a second PMT.

\begin{figure}
\centering{\resizebox{0.95\columnwidth}{!}{\includegraphics{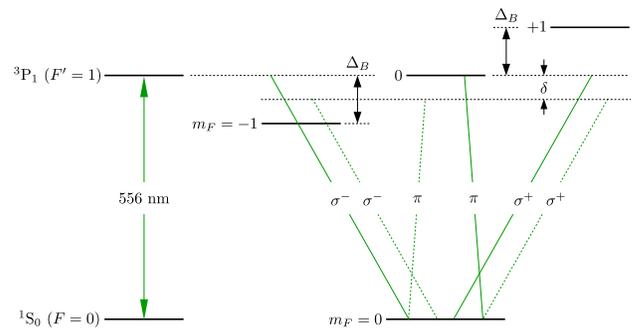}}}
\caption{(Color online) Relevant energy levels of the $\rm {^1S_0} \rightarrow {^3P_1}$ transition in $^{174}$Yb splitting into Zeeman sublevels in the presence of a magnetic field. Sublevels coupled by different polarizations of the pump beam (solid) and the probe beam (dotted) are as shown. The pump beam has a fixed frequency at the $0 \rightarrow 0$ transition. The probe beam is scanned in frequency with a detuning $\delta$.}
 \label{Yblevels556}
\end{figure}

The experiments are done with and without a magnetic field. For this, we wind three pairs of Helmholtz coils on the outside of the vacuum chamber along the three orthogonal directions. For the experiments in zero field, the currents in the three pairs are adjusted to get the most symmetric spectrum, which ensures that the Earth's field and any other stray fields are nullified. The experiments with nonzero field are done by increasing the current in the pair of coils in the direction of the laser beams---the field in this direction can reach up to a maximum of about 650 mG, as measured with a three-axis fluxgate magnetometer. Since the laser beams are linearly polarized and we collect fluorescence in the perpendicular direction, we have to consider transitions coupled by $\sigma^+$, $\sigma^-$, and $\pi$ polarizations, as shown in Fig.\ \ref{Yblevels556}.

\section{Results and discussion}
The ground state of the two-electron atom Yb has the configuration $6s^2$ and the corresponding term symbol $\rm {^1S}_0$. This means that the two electrons are in a singlet state with anti-parallel spins ({\em parasystem} in the language of He), and the state has even parity. The excited state has the configuration $6s \, 6p$ with the term symbol $\rm {^3P}_1$, which implies that the electrons are in a triplet state with parallel spins ({\em orthosystem}), and the state has odd parity. Thus the $\rm {^1S_0} \rightarrow {^3P_1}$ transition, called an intercombination line, is forbidden under L-S coupling (although allowed by parity). But, in a heavy atom like Yb, J-J coupling makes the transition \textit{weakly} allowed---one with a long lifetime of 850 ns (corresponding to a natural linewidth $\Gamma / 2\pi = 185$ kHz).

The experiments are done using the even isotope $^{174}$Yb, which has nuclear spin $I=0$, and hence no hyperfine structure in either the ground state or the excited state. As a consequence, $F=0$ and $F'=1$, as shown in Fig.\ \ref{Yblevels556}. In the presence of a magnetic field $B$, the ground level remains unsplit, while the upper level splits into three sublevels.  The splitting is given by $\Delta_B = g_F \mu_B B$, where $g_F$ is the Land\'e $g$ factor of the level, and $\mu_B$ is the Bohr magneton. For the $F'=1$ level of the $\rm {^3P}_1$ state in $^{174}$Yb, $\Delta_B = 2.1$ MHz/G.

\begin{figure}
\centering{\resizebox{0.95\columnwidth}{!}{\includegraphics{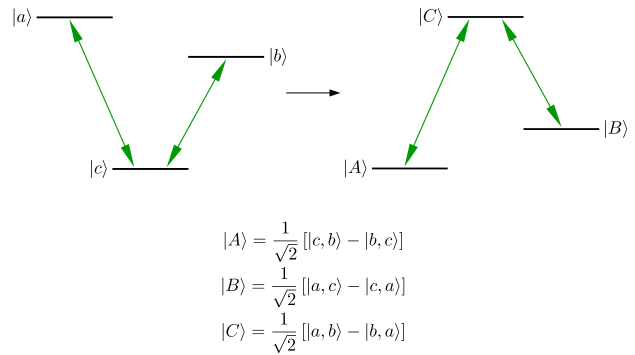}}}
\caption{(Color online) Single-electron states (lower case) in V-type configuration getting transformed to two-electron states (upper case) in $\Lambda$-type configuration. The two-electron states are given below in terms of the one-electron states, with $ \ket{i,j} = \ket{i} \otimes \ket{j} $.}
 \label{vtolambda}
\end{figure}

As has been shown theoretically in Ref.\ \cite{MCR01}, it is possible to observe CPT in such a two-electron atom, even though the levels form a V-type system. The main condition for this to happen is that the two electrons have parallel spins, which is satisfied in the $\rm {^3P}_1$ state of Yb. Since the two electrons have aligned spins, the spin term is symmetric, and the antisymmetry of the total wavefunction comes from the spatial part of the wavefunction. This, combined with the electron-electron interaction, transforms the V-type into a $\Lambda$-type system. The transformation and the resulting two-electron states in terms of the one-electron states are shown in Fig.\ \ref{vtolambda}. The dark state that is formed involves one electron in a superposition of the upper states, while the other electron lies in the lower energy state. Thus a hole appears in the empty upper state, which remains trapped and cannot relax to any other state.

Fig.\ \ref{Yblevels556} shows that six V-type systems are formed, which are listed in Table \ref{cpt}. The last column of the table gives the change in $m_{F'}$ for the pump and probe transitions. All six systems transform into $\Lambda$-type single-electron systems, and result in CPT resonances. 

\begin{table}
\centering{}
\caption{CPT resonances.}
\begin{tabular}{|c|cc|c|c|}
\hline
 & \multicolumn{2}{|c|}{Polarization} & Sublevels &  \\
No. & pump & probe & $m_{F'} \xleftarrow{\rm pump} m_F \xrightarrow{\rm probe} m_{F'}$ & $\Delta m_{F'} $ \\ 
\hline 
1  & $\sigma^- $ & $ \pi$  & $ -1 \leftarrow 0 \rightarrow 0 $ &  $-1$ \\ 
2  & $\pi $ & $ \sigma^+$  & $ 0 \leftarrow 0 \rightarrow +1 $ &  $-1$ \\ 
3  & $\sigma^+ $ & $ \pi$  & $ +1 \leftarrow 0 \rightarrow 0 $ &  $+1$ \\ 
4  & $\pi $ & $ \sigma^-$  & $ 0 \leftarrow 0 \rightarrow -1 $ &  $+1$ \\ 
5  & $\sigma^- $ & $ \sigma^+$  & $ -1 \leftarrow 0 \rightarrow +1 $ &  $-2$ \\ 
6  & $\sigma^+ $ & $ \sigma^-$  & $ +1 \leftarrow 0 \rightarrow -1 $ &  $+2$ \\ 
\hline 
\end{tabular}
\label{cpt}
\end{table}

As mentioned earlier, we have to consider the related phenomenon of CPO, which requires the pump and probe beams to be on the same two levels.  This will happen when the two polarizations are the same, i.e.\ for the three cases listed in Table \ref{cpo}. As expected, $\Delta m_{F'}$ is $0$ for all three cases.

\begin{table}
\centering{}
\caption{CPO resonances.}
\begin{tabular}{|c|cc|c|c|}
\hline
 & \multicolumn{2}{|c|}{Polarization} & Sublevels &  \\
No. & pump & probe & $m_{F'} \xleftarrow{\rm pump} m_F \xrightarrow{\rm probe} m_{F'}$ & $\Delta m_{F'} $ \\ 
\hline 
1  & $\sigma^- $ & $ \sigma^-$  & $ -1 \leftarrow 0 \rightarrow -1 $ &  $0$ \\ 
2  & $\pi $ & $ \pi$  & $ 0 \leftarrow 0 \rightarrow 0 $ &  $0$ \\ 
3  & $\sigma^+ $ & $ \sigma^+$  & $ +1 \leftarrow 0 \rightarrow +1 $ &  $0$ \\
\hline 
\end{tabular}
\label{cpo}
\end{table}

\begin{figure}
\centering{\resizebox{0.95\columnwidth}{!}{\includegraphics{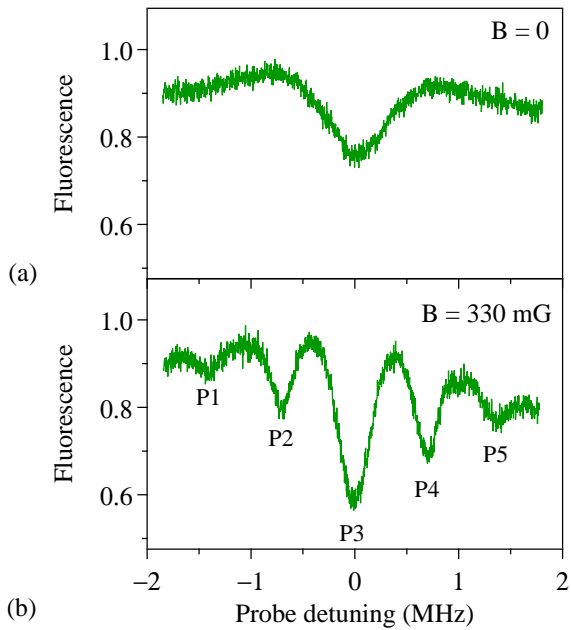}}}
\caption{(Color online) CPT and CPO spectra. The fluorescence dips should appear on a flat background---the slight variation is an artifact and is due to intensity variation of the probe beam. (a) A single dip comprising both CPT and CPO when the levels are degenerate, i.e.\ in the absence of a magnetic field. (b) In the presence of a field, the spectrum splits into 5 dips---the unshifted one P3 corresponds to CPO; while the shifted ones P1, P2, P4, and P5 correspond to CPT. See text for details. }
 \label{noBandB}
\end{figure}

We first consider the case when there is no field, and all the sublevels are degenerate. This means that the six CPT resonances and the three CPO resonances all collapse into a single dip. This is shown in the Fig.\ \ref{noBandB}(a). The single dip is a convolution of all nine dips, and has a linewidth of 800 kHz.

Next, consider what happens when we apply a magnetic field. The CPT resonances will occur when the relative detuning matches the Raman resonance condition:
\[
\Delta_c-\Delta_p = \Delta m_{F'} \, \Delta_B
\]
From Table \ref{cpt}, we can see that the CPT resonances will shift: 1 and 2 to a detuning of $-\Delta_B$; 3 and 4 to a detuning of $+\Delta_B$; 5 to a detuning of $-2 \Delta_B$, and 6 to a detuning of $+2 \Delta_B$. On the other hand, the CPO resonances listed in Table \ref{cpo} will not shift since $\Delta m_{F'} = 0$ for all cases.

This is exactly what is seen in Fig.\ \ref{noBandB}(b). In the presence of a field, the spectrum splits into 5 dips---P1 corresponds to CPT 5; P2 corresponds to CPT 1 and 2; P3 corresponds to CPO 1, 2, and 3; P4 corresponds to CPT 3 and 4; and P5 corresponds to CPT 6. The linewidth of P3 (the unshifted CPO resonance) is 430 kHz, while P2 and P4 (the shifted CPT resonances) have a linewidth of 310 kHz. P1 and P5 do not have enough signal-to-noise ratio to measure the linewidth unambiguously, but it is of the same order of magnitude as P2 and P4. 

The experimental spectra in Fig.\ \ref{noBandB} show two features that require further discussion. One is a background variation of the signal level; this is due to intensity variation as the laser is scanned, which we have verified by using beams of different intensities. The second is that peak P3 is stronger than the adjacent peaks P2 and P4. This is because P3 is a convolution of three resonances, whereas P2 and P4 correspond to two resonances each. This is also the reason for the larger linewidth of P3 compared to the linewidths of P2 and P4. Note that a single CPO resonance will have a linewidth close to the 185 kHz natural linewidth of the $\rm {^3P}_1$ state because that is the excited level involved in the two-level transition. Similarly, the linewidth of a single CPT resonance will be limited by the natural linewidth of the $\rm {^3P}_1$ state because it is involved in the formation of the dark state. But the observed CPT linewidth (from peaks P2 and P4) is slightly larger than this value. This is mainly due to power broadening---the intensity in each beam of 0.41 mW/cm$^2$ is about 3 times the saturation intensity, and needs to be this high in order to get a good signal from the weak intercombination line. 

Transit time can also limit the CPT linewidth in beam experiments. In our case, the transit time is 16.6 \textmu s for an atom traveling with a typical velocity of 300 m/s (the typical velocity for Yb atoms emanating from a source at 400$^\circ$C), going across a beam of size 5 mm. This corresponds to a linewidth of 9.5 kHz, which is much smaller than the observed linewidth. However, to further confirm that transit time is not an issue, we have repeated the experiment with beam sizes that are smaller (3 mm) and bigger (7 mm). There is no measurable difference in the linewidth in the three cases, showing that transit time is not a limitation in our experiment.

\begin{figure}
\centering{\resizebox{0.99\columnwidth}{!}{\includegraphics{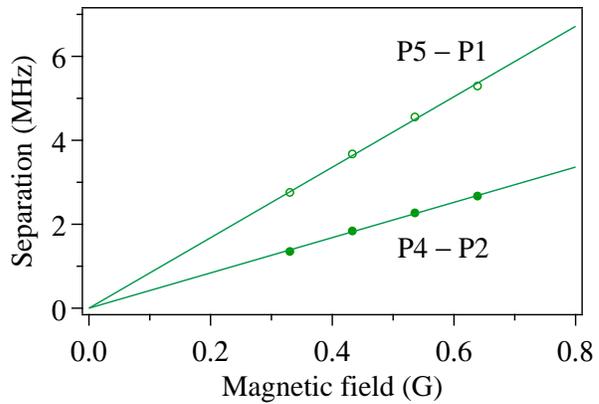}}}
\caption{(Color online) Peak separation vs magnetic field for the two sets of CPT peaks. The error bars (not shown) are about the same size as the symbols. The solid lines are linear fits---they pass through the origin and differ in slope by a factor of 2, as expected.}
 \label{separation}
\end{figure}

In order to verify that the shifted peaks are indeed CPT like, we have studied the separation of the peaks as a function of the applied magnetic field. The experimental handle to change the field in a systematic manner is to vary the current in the coils along the laser beams. The results are shown in Fig.\ \ref{separation}. The symbols represent the measured separation, while the solid lines represent linear fits. As expected from the shifts listed in Table \ref{cpt}, the slope of the separation between peaks P1 and P5 is twice that between P2 and P4. This confirms that the peaks are CPT resonances, and shows that the size of the shift depends on the external magnetic field. This is how CPT resonances can be used for sensitive magnetometry.

\section{Conclusion}

In summary, we have observed CPT resonances in a $F=0 \rightarrow F'=1 $ transition, where the magnetic sublevels form a V-type system. CPT is not possible for such a transition according to one-electron theory because the V-type system does not allow the formation of a long-lived dark state. But CPT is possible in the kind of two-electron atom used here---Yb---because the interaction of the two electrons transforms the V-type into a $\Lambda$-type system, which allows the formation of a long-lived dark state.

Since the experiments are done on a degenerate two-level system, we have to apply a magnetic field to lift the degeneracy. The single fluorescence dip in the absence of a field transforms into five dips in the presence of the field. The unshifted central dip corresponds to CPO, while the four shifted ones are due to CPT. The CPT resonances have a linewidth of about 300 kHz, which is clearly limited by the 185 kHz natural linewidth of the upper state that is involved in forming the dark state. This linewidth is larger than the typical CPT linewidths seen in one-electron atoms, but is still much smaller than the linewidth for E1-allowed transitions (about 10 MHz) because the transition used here is a weakly allowed intercombination line.

\begin{acknowledgments}
This work was supported by the Department of Science and Technology, India. A.K.S.\ acknowledges financial support from the Council of Scientific and Industrial Research, India.
\end{acknowledgments}


\end{document}